\documentclass[a4paper]{article}
\begin{document}
\title{ Relation between Wigner functions and the source functions used in the
description of Bose-Einstein correlations in multiple particle production.}
\author{K.Zalewski\thanks{Partially supported by the KBN grant
2P03B09322}
\\ M.Smoluchowski Institute of Physics
\\ Jagellonian University, Cracow
\\ and\\ Institute of Nuclear Physics, Cracow}

\maketitle
\begin{abstract}
Relations between Wigner functions and the source functions used in models of
Bose-Einstein correlations in multiple particle production are derived and discussed.
These relations are model dependent. In particular it is important whether the
particles are emitted simultaneously and if not, whether the production amplitudes
corresponding to different moments of time can interfere with each other.
\end{abstract}

\section{Introduction}

When discussing Bose-Einstein correlations among identical pions\footnote{Here and in
the following we call for definiteness the identical bosons pions the results,
however, are valid for any identical bosons} produced in multiple particle production
processes, one often uses the source function $S(X,K)$ (cf. e.g. \cite{WIH} and
references quoted there) related to the single particle density matrix in the momentum
representation by the formula\footnote{Often an
equivalent formula with the density matrix replaced by the average \\
$<\hat{a}_p^{\dag}\hat{a}_{p'}> = \rho({\bf p',p})$ is used. Also some authors include
on the left hand side a factor $\sqrt{EE'}$. Since this factor does not affect our
argument and is easy to introduce at any stage, we skip it.}

\begin{equation}\label{source}
\rho(\textbf{p},\textbf{p}') = \int d^4X e^{iqX} S(X,K),
\end{equation}
where $q = p - p'$ and $K = (p + p')/2$ are four-vectors constructed from the on-shell
single particle four momenta $p$ and $p'$. The four components of $X$ are integration
variables and, therefore, there is much freedom in their interpretation. It is usual,
however, to identify $S(X,K)$ with the space-time distribution of the sources
producing the pions with momentum $K$ and then $X$ is interpreted as the space-time
position of the source. A source functions combines conveniently the information
and/or prejudice about the space-time ($X$) distribution of the sources and about the
momentum distribution of the produced particles. It yields the density matrix, which
can be used to get the (two- as well as more particle) correlation function. Comparing
the predicted correlation functions with experiment one can improve the model used to
find $S(X,K)$, fix its parameters etc. Note that the density matrix on the left hand
side of formula (\ref{source}) does not depend on time. Since after freezeout the
particles propagate freely (we do not discuss the final state interactions here), this
means that for times after the freeze out period the elements of the density matrix
should be interpreted as the matrix elements of the time dependent (Schr\"odinger
picture) density operator between the time dependent states $|\textbf{p},t\rangle =
e^{iE(p)t}|\textbf{p}\rangle$, or equivalently as the matrix element of the time
independent (interaction picture) density operator between the time independent states
$|\textbf{p}\rangle$.

Formula (\ref{source}) looks similar to the formula relating the Wigner function
$W(\textbf{X},\textbf{K})$ (cf. e.g. \cite{HCS} and references contained there) to the
density matrix:

\begin{equation}\label{wigner}
\rho(\textbf{p},\textbf{p}',t) = \int d^3X e^{-i\textbf{q}\cdot \textbf{X}}
W(\textbf{X},\textbf{K},t).
\end{equation}
In fact the building block of the source function was originally a Wigner function
(cf. \cite{PRA1} formula (7)). Because of the similarity between formulae
(\ref{source}) and (\ref{wigner}), the source function is often referred to as a
Wigner function, a kind of Wigner function, a pseudo-Wigner function etc. (cf. e.g.
ref. \cite{CHH}, from which all these names have been taken). A discussion of the
actual relation between the source function and the Wigner functions, however, seems
to be missing in the literature. Let us begin with some general remarks.

Formula (\ref{wigner}) contains the time dependent density matrix, i.e. the time
dependent density operator in the representation of the time independent states
$|\textbf{p}\rangle$ . Thus

\begin{equation}\label{rhoodt}
\rho(\textbf{p},\textbf{p}',t) = e^{-iq_0t}\rho(\textbf{p},\textbf{p}').
\end{equation}
Relation (\ref{source}) follows from (\ref{wigner}), if we put

\begin{equation}\label{trivia}
S(X,K) = W({\bf X,K},t_0)\delta(X_0-t_0).
\end{equation}
This choice can be always made and has a simple physical interpretation. Choose a time
$t_0$ after freezeout, when all the pions are already present, but none have been
registered yet. The state at $t_0$ yields the initial condition from which any
distribution at registration can be calculated. Assuming that all the pions appeared
simultaneously at $t_0$ may be poor physics, but this does not change the fact that
the predictions at registration time are correct. There is an infinite choice of other
source functions, which also satisfy relation (\ref{source}). Since this relation is
the only link between a source function and physical reality, there can be no
objection of principle against the choice (\ref{trivia}). From the practical point of
view, however, another source function, which can be calculated from some model, may
be more convenient and, of course, it is just as good, if it reproduces the density
matrix equally well. Note that the identity of functions in (\ref{trivia}) does not
mean that their argument have the same physical interpretation. In fact, on the left
hand side $X$ and $K$ denote the position of the source in space-time and the
four-momentum of the pion, while on the right hand side $\textbf{X} =
\frac{1}{2}(\textbf{x} + \textbf{x}')$ and $\textbf{K} = \frac{1}{2}(\textbf{p} +
\textbf{p}')$ are defined by reference to the arguments of the density matrices of the
pion in the coordinate representation and in the  momentum representation
respectively, while $t_0$ is the time. It is often reasonable to assume that
$\textbf{X}$ and $\textbf{K}$ on both sides are good approximations to the position at
$t = t_0$ and momentum for $t \geq t_0$ of the pion, but this is certainly not always
the case.

Let us conclude this section with some more remarks. If one assumes that all the
particles were created simultaneously at some common freezeout time $t = t_0$, then a
"realistic" source function should be proportional to $\delta(X_0 - t_0)$ and the
proportionality coefficient is the Wigner function as in (\ref{trivia}). The source
function cannot be equal to a Wigner function, because the dimensions are different.
The delta function in formula (\ref{trivia}) brings into the dimension the necessary
factor 1/time. What could be equal to the source function is what we will refer to as
the differential Wigner function defined by the relation

\begin{equation}\label{wigden}
  \tilde{W}(\textbf{X},\textbf{K},t_0) = W_{dt_0}(\textbf{X},\textbf{K},t_0)/dt_0.
\end{equation}
Here $W_{dt_0}$ is the Wigner function at time $t > t_0$ for the particles from
sources created in the time interval $dt_0$ around time $t_0$. Note that we are using
the Wigner functions normalized to the numbers of particles and not to one. When the
amplitudes of the pions produced at different times add incoherently, formula
(\ref{wigner}) can be rewritten in terms of the differential Wigner function
$\tilde{W}$ as:

\begin{equation}\label{wigtin}
  \rho(p,p') = \int d^4X e^{iqX}\tilde{W}(X,\textbf{K}),
\end{equation}
where the fourth component of the vector $X = (\textbf{X},X_0)$ is $t_0$. Comparing
this formula with formula (\ref{source}) we find a particular solution for the source
function

\begin{equation}\label{wisoti}
S(X,K) = \tilde{W}(X,\textbf{K}).
\end{equation}
We stress that this is only one out of the infinity of source functions, which when
substituted into formula (\ref{source}) yield the correct density matrix and that for
the sources incoherence in the creation time has been assumed.

\section{GGLP models}

In the seminal paper of G. Goldhaber, S. Goldhaber, W. Lee and A. Pais \cite{GGL} the
assumption was made that pion production amplitudes at different space points are
incoherent. In this simple case there is no need to distinguish between the sources
and the pions at $t = t_0$. A natural generalization (cf. e.g. \cite{KOP1}) was to
assume more generally that pions production amplitudes at different space-time points
are incoherent. Further we refer to such models as GGLP models.

Let us begin with the simpler case, when all the pions appear simultaneously at
$t=t_0$. For $t > t_0$ they are assumed to propagate freely. The time independent
density matrix for $t > t_0$ is:

\begin{equation} \label{GGLP}
\rho(\textbf{p},\textbf{p}') = \int d^3x \langle \textbf{p},t|\textbf{x},t;t_0\rangle
\rho(\textbf{x},t_0)\langle \textbf{x},t;t_0 |\textbf{p}'t\rangle.
\end{equation}
Here $|\textbf{x},t;t_0\rangle = \exp[-i(t-t_0)\hat{H_0}]|{\bf x}\rangle$ is the time
dependent state, evolving according to the free particle Hamiltonian $\hat{H_0}$ from
the initial state $|\textbf{x}\rangle$, which corresponds to a pion localized at ${\bf
x}$ at time $t_0$. One could also say that in space-time the position of the source is
$(\textbf{x},t_0)$. Let us note for further use the formula

\begin{equation}\label{inclt0}
  \langle \textbf{p},t|\textbf{x},t;t_0\rangle =
  e^{iE(p)t_0}\langle \textbf{p}|\textbf{x} \rangle,
\end{equation}
where $E(p) = \sqrt{\textbf{p}^2 + m_\pi^2}$. Formula (\ref{GGLP}) is perfectly
acceptable from the point of view of quantum mechanics, in spite of the fact that
according to the picture described in ref. \cite{GGL} a pion with momentum ${\bf p}$
is produced at point $\textbf{x}$, which would contradict the uncertainty principle.
Since the density matrix is diagonal in $\textbf{x}$ (in the $\textbf{x}$, which
labels the incoherent states, not in position at time $t$) and $t = t' = t_0$ , we can
identify the components of the vector $\textbf{x}$ with $X_1,X_2,X_3$ and the
parameter $t_0$ with $X_0$. The matrix element (\ref{GGLP}) can be rewritten as

\begin{equation} \label{GGLrho}
\rho(\textbf{p},\textbf{p}') = \int \frac{d^3X}{(2\pi)^3} \exp(iqX) \rho(X).
\end{equation}

Using (\ref{rhoodt}) and comparing with (\ref{wigner}) this yields

\begin{equation}\label{GGLPSr}
W(X,\textbf{K}) =\frac{\rho(X)}{(2\pi)^3}.
\end{equation}
The source function is given by formula (\ref{trivia}), because by assumption all the
particle production happens at $t = t_0$. The components of $X$ are the three
components of the position vector of the point where the particle was produced and its
time of production (freezeout time). Since after freezeout the momentum vector of each
particle is conserved separately, the momenta entering $\textbf{K}$ can be taken at
any time not earlier than $t_0$. The actual time of observation does not occur in the
formula at all.

Note that for this class of models the Wigner function does not depend on momenta.
With the present interpretation this would lead to disaster when comparing with
experiment. The Goldhabers Lee and Pais \cite{GGL} proceeded differently. They were
interested in $n$-pion exclusive states for $n = 4,5,6$ with no more than two pions of
a given charge. Using matrix (\ref{GGLP}) they constructed the probability
distributions $P(p_1,\ldots,p_n)$. These were interpreted as momentum distributions
only after the sets of momenta contradicting energy and/or momentum conservation were
projected out. In practice they interpreted the $P$-s as integrands of the
corresponding phase space integrals. This could not be generalized to inclusive
processes, led to rather complicated integrations and gave results, which at that time
were considered acceptable, but which soon were shown to be in violent disagreement
with experiment \cite{CZS}. Most of these difficulties were overcome by Kopylov,
Podgoretsky and coworkers as described in the next section.

Let us consider now the more general case, when the pions are produced at different
moments of time, but there is no interference between the amplitudes corresponding to
production at different moments.  Then at times larger than the largest freezeout time
the density matrix is given by an integral over the freezeout time $t_0$ of the
(differential version of the)  density matrix (\ref{GGLrho}) corresponding to
particles produced at given time $t_0$:

\begin{equation}\label{inctim}
\rho({\bf p,p'}) = \int \frac{d^4X}{(2\pi)^3} \exp(iqX) \tilde{\rho}(X).
\end{equation}
Note that in order to keep the dimensions right $\rho(X)$, which is a density in
space, has been replaced by $\tilde{\rho}(X)$ which is a density in space time. In
fact, $\rho(X)$ are diagonal elements of the time independent single pion density
matrix in the $|\textbf{x},t;t_0\rangle$ representation, while $\tilde{\rho}(X)$ are
the diagonal elements of the corresponding differential density matrix related to
$\rho(X)$ by a formula analogous to (\ref{wigden}). Comparing with (\ref{source}) it
is seen that one solution for $S(X,K)$ is

\begin{equation}\label{sointi}
  S(X,K) = \frac{\tilde{\rho}(X)}{(2\pi)^3}.
\end{equation}
Since there is incoherence in time, the relation between the source function and
Wigner's function is $S(X,K) = \tilde{W}(X,\textbf{K})$ (\ref{wisoti}).

This seems to be the perfect situation: the source function reproduces exactly the
distribution of (incoherent) sources in space-time and coincides with the differential
Wigner function. However, (\ref{sointi}) is only one of the infinity of source
functions, which yield the same density matrix. A quite different one can be obtained
e.g. by calculating the Wigner function corresponding to the density matrix
$\rho(\textbf{p}_1,\textbf{p}_2,t)$ and substituting it into formula (\ref{trivia}).
We can conclude only that for this class of (unrealistic) models there is an
acceptable source function, which can be interpreted as a differential Wigner
function. Therefore, it can be argued that this particular source function is easier
to find than others.

\section{Including coherence in space}

An important extension of the GGLP approach is to assume that at any given creation
time there may be coherence between the amplitudes of the pions produced at different
points in space, though pions produced at different times $t_0$ do not interfere.
Models of this type, also in the more general version with coherence in time, have
been introduced by Kopylov and Podgoretsky \cite{KOP2}. In terms of sources this may
be quite complicated. For instance, in the model described by Kopylov and Podgoretsky
\cite{KOP2} all the sources are produced simultaneously and live for some time. The
averaging over the energy of each source (not of the pion!), however, kills the
interference between the amplitudes for the production of the pion by the same source
at different moments of time. Therefore one can replace each original source by a set
of sources distributed in time and producing the pions instantly. These are not quite
satisfactory models, because it is difficult to explain why the creation amplitudes
from neighboring time moments should not interfere, while those from neighboring space
points do. Moreover, relativistically the creation processes which are at different
times in one Lorentz frame may be simultaneous in another Lorentz frame. As seen from
formula (\ref{trivia}), however, even in the simplest case of simultaneous production,
such models can reproduce correctly any single particle density matrix and
consequently any single particle momentum distribution without doing phase space
integrations. Thus they can be applied to describe inclusive processes.

Including interference in space and summing without interference over time corresponds
to the following generalization of formula (\ref{GGLP})

\begin{equation} \label{xcoher}
\rho(\textbf{p},\textbf{p}') = \int dt_0\int d^3x \int d^3x' \langle \textbf{p},t|
\textbf{x},t;t_0 \rangle\tilde{\rho}( \textbf{x}, \textbf{x}', t_0)\langle
\textbf{x}', t; t_0|\textbf{p}', t\rangle.
\end{equation}

There are many equivalent ways of rewriting this relation. For instance, one may use a
representation $|\alpha\rangle$, where the (differential) density matrix
$\tilde{\rho}(\textbf{x},\textbf{x}',t_0)$ is diagonal. Then (\ref{xcoher}) gets
replaced by

\begin{equation}\label{acoher}
\rho(\textbf{p},\textbf{p}') = \int dt_0\int d\alpha e^{iq_0t_0}\langle
\textbf{p}|\alpha\rangle \tilde{\rho}(\alpha;t_0) \langle \alpha|\textbf{p}'\rangle.
\end{equation}
It is instructive to compare this model, to the GGLP models. At $t_0$ the localized
states $|\textbf{x}\rangle = |\textbf{x},t_0;t_0\rangle$ have been replaced by the
states $|\alpha \rangle = |\alpha, t_0;t_0\rangle$. The plane waves $\langle
\textbf{p}|\textbf{x}\rangle$ have been replaced by the functions $\langle
\textbf{p}|\alpha\rangle$, which are known under a variety of names: as waved packets,
as sources, or as currents. The source function for such models was introduced by
Pratt \cite{PRA1}. His statement that\footnote{This has been changed to our notation.
Pratt has written $g,\vec{p},x$\ where we have written $S,\textbf{K},X$} "
$S(X,\textbf{K})$ can be identified as the probability of emitting a pion of momentum
$\textbf{K}$ from space-time point X" is, however, only an approximation.

Kopylov and Podgoretsky \cite{KOP2} assumed that $(\alpha,t_0)$ is a point in
space-time, so that formula (\ref{acoher}) can be rewritten as

\begin{equation}\label{kpwapa}
\rho(\textbf{p},\textbf{p}') =\int d^4x_0 \langle
\textbf{p}|\psi_{x_0}\rangle\tilde{\rho}(x_0) \langle \psi_{x_0}|\textbf{p}'\rangle
e^{iq_0t_0}.
\end{equation}
This approach yields an alternative method of describing the distribution of pions in
phase space. In the integrand $\langle \textbf{p}|\psi_{x_0}\rangle$ is the
probability amplitude for finding a pion with momentum $\textbf{p}$ produced by a
source labelled $x_0$. When the states $|\psi_{x_0}\rangle$ correspond to particle
well localized in space, $\textbf{p}$ and $\textbf{x}_0$ give a reasonably good
description of the position of the pion in phase space. One could object that this is
only a rough description, but the same is true for the Wigner function: $\textbf{K}$
and $\textbf{X}$ give only approximately the momentum and position of the pion. An
exact determination of the pion position in phase space is possible only in classical
physics.

In ref. \cite{KOP2} the states $|\alpha\rangle = |\psi_{x_0}\rangle$ were supposed to
be related by space time translations so that

\begin{eqnarray}\label{kptreq}
  \psi_{x_0}(x) &\equiv & \langle \textbf{x}|\psi_{x_0}\rangle = \psi(x - x_0),\\
  \langle \textbf{p}|\psi_{x_0}\rangle &= & e^{ipx_0}\phi(\textbf{p}),\qquad \phi(\textbf{p}) =
  \langle \textbf{p}|\psi_0\rangle.
\end{eqnarray}
The second formula is, of course, equivalent to the first. Another choice has been
made in the "covariant current ensemble formalism" \cite{KOG}, \cite{GYP}, \cite{PGG},
\cite{PAG}. There each source is labelled by a position in space-time $x_0$ and a
four-momentum $p_0$. Usually $\textbf{x}_0$ and $\textbf{p}_0$ denote the centers of
the wave packet in ordinary space and in momentum space respectively. The time
component of $x_0$ is $t_0$ -- the freeze out time of the wave packet. The time
component of $p_0$ is calculated from the condition $p_0^2 = m^2$, where $m$ can
\cite{PAG}, but does not have to \cite{CHH} be the pion mass. The nice feature of this
parametrization of the wave packets is that one can substitute a classical trajectory
$\textbf{x}_0(t),\textbf{p}_0(t)$ for the source and remain in agreement with the
Heisenberg uncertainty principle for the pions. The choice for the scalar products is

\begin{equation}\label{curgyu}
\langle \textbf{p}|\psi_{x_0,p_0}\rangle = e^{ipx_0}j(\frac{pp_0}{m}).
\end{equation}
Thus, the sources are related by space-time translations and when each of the currents
$j$ is considered in its rest frame where $p_0 = (m,0)$, they  are identical. Formulae
(\ref{kptreq}) and (\ref{curgyu}) have been applied also to relativistically covariant
models cf. e.g. \cite{KOP2}, \cite{PGG}.

Another way of rewriting relation (\ref{xcoher}) is

\begin{equation}\label{xcohxy}
\rho({\bf p,p'}) = \int d^4X \exp(iqX) \int \frac{d^3y}{(2\pi)^3} \exp(-i{\bf K\cdot
\textbf{y}})\tilde{\rho}({\bf x,x'},t_0),
\end{equation}
where  ${\bf X}=({\bf x + x'})/2$, ${\bf y} = {\bf x - x'}$ and $X_0 = t_0$. According
to formula (\ref{source}) a possible choice of the source function is

\begin{equation}\label{intsou}
S(X,K) = \int \frac{d^3y}{(2\pi)^3} \exp(-i\textbf{K}\cdot \textbf{y})
\tilde{\rho}(\textbf{x},\textbf{x}',t_0).
\end{equation}
The differential Wigner function is the Wigner transform of the differential density
matrix as it should. The relation of this particular source function to the Wigner
function is again given by formula (\ref{wisoti}). We conclude that among the
infinitely many source functions which give the same density matrix in the momentum
representation there is one, which can be related to the Wigner function as described.
Note that this source function depends on $\textbf{K}$, but  does not depend on $K_0$.

\section{Including coherence in space and time}

Finally let us consider the case, when neither coherence in space, nor coherence in
time is assumed. Then the density matrix is

\begin{equation}\label{incinc}
\rho(\textbf{p},\textbf{p}') = \int d^4x \int d^4x'\langle
\textbf{p},t|\textbf{x},t;t_0\rangle\tilde{\rho}(x,x') \langle
\textbf{x}',t;t_0'|\textbf{p}',t\rangle,
\end{equation}
where $x = (\textbf{x},t_0)$ and $x' = (\textbf{x}',t_0')$. For an application of a
formula of this type see e.g. ref. \cite{PCZ}, where $\sum_F M_F(x)M_F(x')$ stand for
our $\tilde{\rho}(x,x')$.

Formula (\ref{incinc}) can be rewritten as

\begin{equation}\label{ininxy}
\rho(\textbf{p},\textbf{p}') = \int d^4X \int d^4y \exp[iqX +
iKy]\tilde{\rho}(x,x')/(2\pi)^3,
\end{equation}
where $X = \frac{1}{2}(x + x')$ and $y = x - x'$. Comparison with formula
(\ref{source}) gives as one of the solutions for the source function

\begin{equation}\label{soinin}
S(X,K) = \int \frac{d^4y}{(2\pi)^3} \exp[iKy]\tilde{\rho}(x,x').
\end{equation}
The right hand side is rather remote from what one usually calls a Wigner function.
The differential Wigner function was at least  proportional to a Wigner function,
though only for the particles from sources produced in the time interval $dt_0$ around
the time $t = t_0$. Function (\ref{soinin}) is proportional to the contribution to the
Wigner function from the interference of the production amplitude in the time interval
$dt_0$ around $t = t_0$ with the production amplitude in the time interval $dt_0'$
around $t = t_0'$, integrated over $t_0' - t_0$ at fixed $t_0' + t_0$.

Formula (\ref{incinc}) rewritten in terms of wave packets reads

\begin{equation}\label{wpinin}
\rho(\textbf{p},\textbf{p}') = \int dt_0 \int dt_0'\int d\alpha \langle
\textbf{p},t|\alpha,t;t_0\rangle\tilde{\rho}(\alpha,t_0,t_0') \langle
\alpha,t;t_0'|\textbf{p}',t\rangle.
\end{equation}
Note that the state $|\alpha,t;t_0\rangle$ may, but does not have to, be connected to
the state $|\alpha,t;t_0'\rangle$ by a smooth hamiltonian evolution. The source at
time $t_0$ can be, as well, something quite different from the source at time $t_0'$.
On the other hand, the evolution of $|\alpha, t; t_0\rangle$ in the time $t$ for $t$
later than the freezeout of the particle is the ordinary free particle evolution. The
various models are defined by the choice of $\alpha$, of the states
$|\alpha,t;t_0\rangle$ and of the weight function $\tilde{\rho}(\alpha,t_0,t_0')$.
Examples can be found in refs. \cite{KOP2}, \cite{PGG}, \cite{CHH}.

\section{Conclusions}

A source function cannot be equal to a Wigner function, because they have different
dimensions. Moreover, for a given state of the system its Wigner function is well
defined, while its source function is not. The problem is, therefore, to chose some
special source function and try to relate it somehow to a Wigner function.

When all the particles are produced simultaneously and when this is assumed to mean
that the source function is proportional to a delta function in time, the
proportionality coefficient is unambiguously defined as the Wigner function of the
pions at the time of freeze out (\ref{trivia}). When it is assumed that the production
amplitudes at different times add incoherently, one can use the source function
proportional to the differential Wigner function as given in (\ref{sointi}). When the
production amplitudes at different times interfere, a source function can be related
to a piece of the Wigner function as given by formula (\ref{soinin}) and explained
below this formula. In this case, however, the use of wave packets (or sources, or
currents) may be a more natural way to analyze the phase space distribution of pions.

If, as is often done, a model is defined by postulating the source function (cf.
\cite{WIH} for examples), the question about the relation of this source function to a
Wigner function cannot be answered without making an assumption about the coherence or
incoherence of the amplitudes for pion production at different times. When the source
function is proportional to a delta function in time, one can relate it to a Wigner
function by relation (\ref{trivia}). When it does not depend on $K_0$, one can use
formula (\ref{wisoti}). In the later case there is no guarantee that the production
process was such as the differential Wigner functions suggest.


\begin{thebibliography} {99}
\bibitem{WIH}U.A. Wiedemann and U. Heinz, {\it Phys. Rep.} {\bf 319}(1999)145.
\bibitem{HCS}M. Hillery, R.F O'Connel, M. Scully and E.P. Wigner, {\it Phys
Rep.} {\bf 106}(1984)121.
\bibitem{PRA1}S. Pratt, \textit{Phys. Rev. Letters} \textbf{53}(1984)1219.
\bibitem{CHH}S. Champan and U. Heinz, {\it Phys. Letters} {\bf B340}(1994)250.
\bibitem{GGL}G. Goldhaber, S. Goldhaber, W. Lee and A. Pais, {\it Phys. Rev.}
{\bf 120}(1960)300.
\bibitem{KOP1}G.I. Kopylov and M.I. Podgoretsky, {\it Yad. Phys.} {\bf
15}(1972)103.
\bibitem{CZS}O. Czy\.{z}ewski and M. Szeptycka, \textit{Phys. Letters} \textbf{25B}(1967)482.
\bibitem{KOP2}G.I. Kopylov and M.I. Podgoretsky, {\it Yad. Phys.} {\bf
19}(1974)434.
\bibitem{PCZ}S. Pratt, T. Cs\"org\"o and J. Zim\'anyi, \textit{Phys. Rev.}
\textbf{C42}(1990)2646.
\bibitem{KOG}K. Kolehmeinen and M. Gyulassy, {\it Phys. Letters} {\bf
B180}(1986)203.
\bibitem{GYP}M. Gyulassy and S.S. Padula, \textit{Phys. Letters} \textbf{B217}(1988)181.
\bibitem{PGG}S.S. Padula, M. Gyulassy and S. Gavin, \textit{Nucl. Phys.} \textbf{B329}(1990)357.
\bibitem{PAG}S.S. Padula and M. Gyulassy, {\it Nucl. Phys.} {\bf
B339}(1990)378.
\end{thebibliography}
\end{document}